\documentclass[journal]{IEEEtran}

\usepackage{cite}
\usepackage{acronym}
\usepackage{graphicx}
\usepackage{tabularx}
\usepackage{enumerate}
\usepackage{color}
\usepackage{amsmath}
\usepackage{paralist}
\usepackage{threeparttable}
\usepackage{booktabs}
\usepackage{colortbl,multirow,hhline}
\usepackage{boldline}
\usepackage{colortbl}
\usepackage{array}
\usepackage{adjustbox}
\usepackage{multirow,tabu}
\usepackage{algorithm,algpseudocode}
\usepackage{tikz,pgfplots}
\usetikzlibrary{arrows}
\usepackage{pgfplotstable}
\usepgfplotslibrary{units,groupplots}
\usepackage{wrapfig}
\usepackage{varwidth,xcolor}
\usepackage{subfigure}
\usepackage{fancyhdr}

\newcommand{\mytinysize}{\fontsize{6}{7}\selectfont}
\pgfplotsset{
	compat = newest,
	tick label style={font=\sffamily\scriptsize},
	label style={font=\sffamily\scriptsize},
	legend style={font=\sffamily\mytinysize\raggedleft},
	legend cell align=left,
	grid style={dotted,gray}
}

\newcolumntype{?}{!{\vrule width 0.8pt}}

\definecolor{mygray}{RGB}{220,220,220}
\definecolor{skyblue}{RGB}{86,180,233}
\definecolor{bluish-green}{RGB}{0,158,115}
\definecolor{myblue}{RGB}{86,114,178}
\definecolor{vermilion}{RGB}{213,94,0}
\definecolor{reddish-purple}{RGB}{204,121,167}

\newcommand{\features}{11}
\newcommand{\ourtool}[0]{\textsc{Lucid}}

\newlength{\Oldarrayrulewidth}
\newcommand{\Cline}[2]{%
	\noalign{\global\setlength{\Oldarrayrulewidth}{\arrayrulewidth}}%
	\noalign{\global\setlength{\arrayrulewidth}{#1}}\cline{#2}%
	\noalign{\global\setlength{\arrayrulewidth}{\Oldarrayrulewidth}}}

\graphicspath{{artworks/}}
\DeclareGraphicsExtensions{.pdf,.jpeg,.png}
\begin{document}

\title{\ourtool: A Practical, Lightweight Deep Learning Solution for DDoS Attack Detection}

\author{
	R. Doriguzzi-Corin$^\alpha$,
	S. Millar$^\beta$,
	S. Scott-Hayward$^\beta$,
	J. Mart\'inez-del-Rinc\'on$^\beta$,
	D. Siracusa$^\alpha$\\
	
	\small{$^\alpha$ICT, Fondazione Bruno Kessler - Italy}\\
	\small{$^\beta$CSIT, Queen's University Belfast - Northern Ireland}
	
}

\maketitle

\thispagestyle{fancy}
\renewcommand{\headrulewidth}{0pt}
\chead{\scriptsize This is the author's version of an article that has been published in IEEE Transactions on Network and Service Management. Changes were made to\\this version by the publisher prior to publication. The final version of record is available at {\color{blue}{https://doi.org/10.1109/TNSM.2020.2971776}}. \\ The source code associated with this project is available at {\color{blue}{https://github.com/doriguzzi/lucid-ddos}}.}
\cfoot{\scriptsize Copyright (c) 2020 IEEE. Personal use is permitted. For any other purposes, permission must be obtained from the IEEE by emailing pubs-permissions@ieee.org.}

\begin{abstract}
\ac{ddos} attacks are one of the most harmful threats in today's Internet, disrupting the availability of essential services. The challenge of \ac{ddos} detection is the combination of attack approaches coupled with the volume of live traffic to be analysed. In this paper, we present a practical, lightweight deep learning \ac{ddos} detection system called \ourtool, which exploits the properties of \acp{cnn} to classify traffic flows as either malicious or benign. We make four main contributions; (1) an innovative application of a \ac{cnn} to detect \ac{ddos} traffic with low processing overhead, (2) a dataset-agnostic preprocessing mechanism to produce traffic observations for online attack detection, (3) an activation analysis to explain \ourtool{}'s \ac{ddos} classification, and (4) an empirical validation of the solution on a resource-constrained hardware platform. Using the latest datasets, \ourtool\ matches existing state-of-the-art detection accuracy whilst presenting a 40x reduction in processing time, as compared to the state-of-the-art. With our evaluation results, we prove that the proposed approach is suitable for effective \ac{ddos} detection in resource-constrained operational environments. 
\end{abstract}


\maketitle

\acrodef{acl}[ACL]{Access Control List}
\acrodef{ann}[ANN]{Artificial Neural Network}
\acrodef{api}[API]{Application Programming Interface}
\acrodef{bow}[BoW]{Bag-of-Words}
\acrodef{cnn}[CNN]{Convolutional Neural Network}
\acrodef{cpe}[CPE]{Customer Premise Equipment}
\acrodef{dl}[DL]{deep learning}
\acrodef{dlp}[DLP]{Data Loss/Leakage Prevention}
\acrodef{dpi}[DPI]{Deep Packet Inspection}
\acrodef{dos}[DoS]{Denial of Service}
\acrodef{ddos}[DDoS]{Distributed Denial of Service}
\acrodef{ebpf}[eBPF]{extended Berkeley Packet Filter}
\acrodef{ewma}[EWMA]{Exponential Weighted Moving Average}
\acrodef{foss}[FOSS]{Free and Open-Source Software}
\acrodef{fpr}[FPR]{False Positive Rate}
\acrodef{gpu}[GPU]{Graphics Processing Unit}
\acrodef{ha}[HA]{Hardware Appliance}
\acrodef{ids}[IDS]{Intrusion Detection System}
\acrodef{ilp}[ILP]{Integer Linear Programming}
\acrodef{iot}[IoT]{Internet of Things}
\acrodef{isp}[ISP]{Internet Service Provider}
\acrodef{ips}[IPS]{Intrusion Prevention System}
\acrodef{lstm}[LSTM]{Long Short-Term Memory}
\acrodef{mano}[NFV MANO]{NFV Management and Orchestration}
\acrodef{mips}[MIPS]{Millions of Instructions Per Second}
\acrodef{ml}[ML]{Machine Learning}
\acrodef{mlp}[MLP]{Multi-Layer Perceptron}
\acrodef{nat}[NAT]{Network Address Translation}
\acrodef{nic}[NIC]{Network Interface Controller}
\acrodef{nids}[NIDS]{Network Intrusion Detection System}
\acrodef{nf}[NF]{Network Function}
\acrodef{nfv}[NFV]{Network Function Virtualization}
\acrodef{nsc}[NSC]{Network Service Chaining}
\acrodef{of}[OF]{OpenFlow}
\acrodef{os}[OS]{Operating System}
\acrodef{pess}[PESS]{Progressive Embedding of Security Services}
\acrodef{pop}[PoP]{Point of Presence}
\acrodef{ppv}[PPV]{Positive Predictive Value}
\acrodef{ps}[PS]{Port Scanner}
\acrodef{qoe}[QoE]{Quality of Experience}
\acrodef{qos}[QoS]{Quality of Service}
\acrodef{rnn}[RNN]{Recurrent Neural Network}
\acrodef{sdn}[SDN]{Software Defined networking}
\acrodef{sla}[SLA]{Service Level Agreement}
\acrodef{snf}[SNF]{Security Network Function}
\acrodef{svm}[SVM]{Support Vector Machine}
\acrodef{tc}[TC]{Traffic Classifier}
\acrodef{tor}[ToR]{Top of Rack}
\acrodef{tpr}[TPR]{True Positive Rate}
\acrodef{tsp}[TSP]{Telecommunication Service Provider}
\acrodef{unb}[UNB]{University of New Brunswick}
\acrodef{vm}[VM]{Virtual Machine}
\acrodef{vne}[VNE]{Virtual Network Embedding}
\acrodef{vnep}[VNEP]{Virtual Network Embedding Problem}
\acrodef{vnf}[VNF]{Virtual Network Function}
\acrodef{vsnf}[VSNF]{Virtual Security Network Function}
\acrodef{vpn}[VPN]{Virtual Private Network}
\acrodef{xdp}[XDP]{eXpress Data Path}
\acrodef{wan}[WAN]{Wide Area Network}
\acrodef{waf}[WAF]{Web Application Firewall}

\section{Introduction} \label{sec:introduction} 

\ac{ddos} attacks are one of the most harmful threats in today's Internet, disrupting the availability of essential services in production systems and everyday life. Although \ac{ddos} attacks have been known to the network research community since the early 1980s, our network defences against these attacks still prove inadequate. 

In late 2016, the attack on the Domain Name Server (DNS) provider, Dyn, provided a worrying demonstration of the potential disruption from targeted \ac{ddos} attacks \cite{ddos-spotify}. This particular attack leveraged a botnet (Mirai) of unsecured IoT (Internet of Things) devices affecting more than 60 services. At the time, this was the largest \ac{ddos} attack recorded, at 600\,Gbps. This was exceeded in February 2018 with a major \ac{ddos} attack towards Github \cite{memcached}. At its peak, the victim saw incoming traffic at a rate of 1.3\,Tbps. The attackers leveraged a vulnerability present in memcached, a popular database caching tool. In this case, an amplification attack was executed using a spoofed source IP address (the victim IP address). If globally implemented, BCP38 ``Network Ingress Filtering'' \cite{bcp38} could mitigate such an attack by blocking packets with spoofed IP addresses from progressing through the network. However, these two examples illustrate that scale rather than sophistication enables the \ac{ddos} to succeed. 

In recent years, \ac{ddos} attacks have become more difficult to detect due to the many combinations of attack approaches. For example, multi-vector attacks where an attacker uses a combination of multiple protocols for the \ac{ddos} are common. In order to combat the diversity of attack techniques, more nuanced and more robust defence techniques are required. Traditional signature-based intrusion detection systems cannot react to new attacks. Existing statistical anomaly-based detection systems are constrained by the requirement to define thresholds for detection. \acp{nids} using machine learning techniques are being explored to address the limitations of existing solutions. In this category, \ac{dl} systems have been shown to be very effective in discriminating \ac{ddos} traffic from benign traffic by deriving high-level feature representations of the traffic from low-level, granular features of packets \cite{7946998,8482019}. However, many existing \ac{dl}-based approaches described in the scientific literature are too resource-intensive from the training perspective, and lack the pragmatism for real-world deployment. Specifically, current solutions are not designed for online attack detection within the constraints of a live network where detection algorithms must process traffic flows that can be split across multiple capture time windows.

\acfp{cnn}, a specific \ac{dl} technique, have grown in popularity in recent times leading to major innovations in computer vision \cite{resnet, imagenet,sabokrou} and Natural Language Processing \cite{KimConvolutionalNN}, as well as various niche areas such as protein binding prediction \cite{deepbind, DanQ}, machine vibration analysis \cite{janssens} and medical signal processing \cite{sleepCNN}.  Whilst their use is still under-researched in cybersecurity generally, the application of \acp{cnn} has advanced the state-of-the-art in certain specific scenarios such as malware detection \cite{AndroidCNNCODA, MultiModelDLAndroid, wang, yeo}, code analysis \cite{sourcecodecnn}, network traffic analysis \cite{7946998, wu2018novel, potluri2018convolutional, vinayakumar} and intrusion detection in industrial control systems~\cite{aads}.  These successes, combined with the benefits of \ac{cnn} with respect to reduced feature engineering and high detection accuracy, motivate us to employ \acp{cnn} in our work.  

While large \ac{cnn} architectures have been proven to provide state-of-the-art detection rates, less attention has been given to minimise their size while maintaining competent performance in limited resource environments. As observed with the Dyn attack and the Mirai botnet, the opportunity for launching \ac{ddos} attacks from unsecured IoT devices is increasing as we deploy more IoT devices on our networks. This leads to consideration of the placement of the defence mechanism. Mitigation of attacks such as the Mirai and Memcached examples include the use of high-powered appliances with the capacity to absorb volumetric \ac{ddos} attacks. These appliances are located locally at the enterprise or in the Cloud. With the drive towards edge computing to improve service provision, it becomes relevant to consider the ability to both protect against attacks closer to the edge and on resource-constrained devices. Indeed, even without resource restrictions, it is valuable to minimize resource usage for maximum system output.

Combining the requirements for advanced \ac{ddos} detection with the capability of deployment on resource-constrained devices, this paper makes the following contributions:
\begin{itemize}
	\item A \ac{dl}-based \ac{ddos} detection architecture suitable for online resource-constrained environments, which leverages \acp{cnn} to learn the behaviour of \ac{ddos} and benign traffic flows with both low processing overhead and attack detection time. We call our model \ourtool\ (Lightweight, Usable CNN in DDoS Detection). 
	\item A dataset-agnostic preprocessing mechanism that produces traffic observations consistent with those collected in existing online systems, where the detection algorithms must cope with segments of traffic flows collected over pre-defined time windows.
	\item A kernel activation analysis to interpret and explain to which features \ourtool{} attaches importance when making a \ac{ddos} classification.
	\item An empirical validation of \ourtool\ on a resource-constrained hardware platform to demonstrate the applicability of the approach in edge computing scenarios, where devices possess limited computing capabilities.    
\end{itemize}

The remainder of this paper is structured as follows: Sec. \ref{sec:related} reviews and discusses the related work. Sec. \ref{sec:methodology} details the methodology with respect to the network traffic processing and the \ourtool\ \ac{cnn} model architecture. Sec. \ref{sec:setup} describes the experimental setup detailing the datasets and the development of \ourtool\ with the hyper-parameter tuning process. In Sec. \ref{sec:evaluation}, \ourtool\ is evaluated and compared with the state-of-the-art approaches. Sec. \ref{sec:analysis} introduces our kernel activation analysis for explainability of \ourtool{}'s classification process. Sec. \ref{sec:usecase} presents the experiment and results for the \ac{ddos} detection at the edge. Finally, the conclusions are provided in Sec. \ref{sec:conclusions}.

\section{Related work}\label{sec:related}

\ac{ddos} detection and mitigation techniques have been explored by the network research community since the first reported \ac{ddos} attack incident in 1999 \cite{criscuolo2000distributed}. In this section, we review and discuss anomaly-based \ac{ddos} detection techniques categorised by statistical approaches and machine learning approaches, with a specific focus on deep learning techniques.

\subsection{Statistical approaches to \ac{ddos} detection}\label{sec:related-stats}
Measuring statistical properties of network traffic attributes is a common approach to \ac{ddos} detection, and generally involves monitoring the entropy variations of specific packet header fields. By definition, the entropy is a measure of the diversity or the randomness in a data set. Entropy-based \ac{ddos} detection approaches have been proposed in the scientific literature since the early 2000s, based on the assumption that during a volumetric \ac{ddos} attack, the randomness of traffic features is subject to sudden variations. The rationale is that volumetric \ac{ddos} attacks are typically characterised by a huge number of attackers (in the order of hundreds of thousands \cite{8367750}), often utilising compromised devices that send a high volume of traffic to one or more end hosts (the victims). As a result, these attacks usually cause a drop in the distribution of some of the traffic attributes, such as the destination IP address, or an increase in the distribution of other attributes, such as the source IP address. The identification of a \ac{ddos} attack is usually determined by means of thresholds on these distribution indicators.

In one of the first published works using this approach, Feinstein et al. \cite{1194894} proposed a \ac{ddos} detection technique based on the computation of source IP address entropy and Chi-square distribution. The authors observed that the variation in source IP address entropy and chi-square statistics due to fluctuations in legitimate traffic was small, compared to the deviations caused by \ac{ddos} attacks. Similarly, \cite{BOJOVIC201984} combined entropy and volume traffic characteristics to detect volumetric \ac{ddos} attacks, while the authors of \cite{8466805} proposed an entropy-based scoring system based on the destination IP address entropy and dynamic combinations of IP and TCP layer attributes to detect and mitigate DDoS attacks.

A common drawback to these entropy-based techniques is the requirement to select an appropriate detection threshold. Given the variation in traffic type and volume across different networks, it is a challenge to identify the appropriate detection threshold that minimizes false positive and false negative rates in different attack scenarios. One solution is to dynamically adjust the thresholds to auto-adapt to the normal fluctuations of the network traffic, as proposed in \cite{10.1007/978-981-13-2622-6_31,8423699}. 

Importantly, monitoring the distribution of traffic attributes does not provide sufficient information to distinguish between benign and malicious traffic. To address this, some approaches apply a rudimentary threshold on the packet rate \cite{Jun:2014:DAD:2554850.2555109} or traceback techniques \cite{5467062,7345297}. 

An alternative statistical approach is adopted in \cite{8522048}, where Ahmed et al. use packet attributes and traffic flow-level statistics to distinguish between benign and \ac{ddos} traffic. However, this solution may not be suitable for online systems, since some of the flow-level statistics used for the detection e.g. total bytes, number of packets from source to destination and from destination to source, and flow duration, cannot be computed when the traffic features are collected within observation time windows. Approaches based on flow-level statistics have also been proposed in  \cite{7997375,8599821,tr-ids,10.1007/978-3-030-00018-9_15,8666588,homayoun}, among many others. In particular, \cite{tr-ids,10.1007/978-3-030-00018-9_15,8666588,homayoun} use flow-level statistics to feed \acp{cnn} and other \ac{dl} models, as discussed in Sec. \ref{sec:related-dl}. To overcome the limitations of statistical approaches to \ac{ddos} detection, machine learning techniques have been explored.

\subsection{Machine Learning for \ac{ddos} detection}\label{sec:related-ml}
As identified by Sommer and Paxson in \cite{sommer2010outside}, there has been extensive research on the application of machine learning to network anomaly detection. The 2016 Buczak and Guven survey \cite{buczak2016survey} cites the use of \ac{svm}, k-Nearest Neighbour (k-NN), Random Forest, Na\"ive Bayes etc. achieving success for cyber security intrusion detection. However, due to the challenges particular to network intrusion detection, such as high cost of errors, variability in traffic etc., adoption of these solutions in the ``real-world'' has been limited. Over recent years, there has been a gradual increase in availability of realistic network traffic data sets and an increased engagement between data scientists and network researchers to improve model explainability such that more practical \ac{ml} solutions for network attack detection can be developed.
Some of the first application of machine learning techniques specific to \ac{ddos} detection has been for traffic classification. Specifically, to distinguish between benign and malicious traffic, techniques such as extra-trees and multi-layer perceptrons have been applied \cite{Idhammad2018,Singh2016EntropyBasedAL}. 

In consideration of the realistic operation of \ac{ddos} attacks from virtual machines, He et al. \cite{he2017machine} evaluate nine \ac{ml} algorithms to identify their capability to detect the \ac{ddos} from the source side in the cloud. The results are promising with high accuracy (99.7\%) and low false positives ($<$ 0.07\%) for the best performing algorithm; \ac{svm} linear kernel. Although there is no information provided regarding the detection time or the datasets used for the evaluation, the results illustrate the variability in accuracy and performance across the range of \ac{ml} models. This is reflected across the literature e.g. \cite{hoon2018critical,primartha2017anomaly} with the algorithm performance highly dependent on the selected features (and datasets) evaluated. This has motivated the consideration of deep learning for \ac{ddos} detection, which reduces the emphasis on feature engineering.

\subsection{Deep Learning for \ac{ddos} detection}\label{sec:related-dl}
There is a small body of work investigating the application of \ac{dl} to \ac{ddos} detection. For example, in \cite{8343104}, the authors address the problem of threshold setting in entropy-based techniques by combining entropy features with \ac{dl}-based classifiers. The evaluation demonstrates improved performance over the threshold-based approach with higher precision and recall. In \cite{yin2017deep}, a \ac{rnn}-\ac{ids} is compared with a series of previously presented \ac{ml} techniques (e.g. J48, \ac{ann}, Random Forest, and \ac{svm}) applied to the NSL-KDD~\cite{5356528} dataset. The \ac{rnn} technique demonstrates a higher accuracy and detection rate. 

Some \ac{cnn}-based works \cite{tr-ids, 10.1007/978-3-030-00018-9_15,8666588,homayoun}, as identified in Sec. \ref{sec:related-stats}, use flow-level statistics (total bytes, flow duration, total number of flags, etc.) as input to the proposed \ac{dl}-based architectures. In addition, \cite{tr-ids} and \cite{10.1007/978-3-030-00018-9_15} combine the statistical features with packet payloads to train the proposed \acp{ids}.

In \cite{wu2018novel}, Kehe Wu et al. present an \ac{ids} based on \ac{cnn} for multi-class traffic classification. The proposed neural network model has been validated with flow-level features from the NSL-KDD dataset encoded into 11x11 arrays. Evaluation results show that the proposed model performs well compared to complex models with 20 times more trainable parameters. A similar approach is taken by the authors of \cite{potluri2018convolutional}, where the \ac{cnn}-based \ac{ids} is validated over datasets NSL-KDD and UNSW-NB-15~\cite{7348942}. In \cite{kwon2018empirical}, the authors study the application of \acp{cnn} to \ac{ids} by comparing a series of architectures (shallow, moderate, and deep, to reflect the number of convolution and pooling layers) across 3 traffic datasets; NSL-KDD, Kyoto Honeypot \cite{song2006description}, and MAWILab \cite{callegari2016statistical}. In the results presented, the shallow \ac{cnn} model with a single convolution layer and single max. pooling layer performed best. However, there is significant variance in the detection accuracy results across the datasets, which indicates instability in the model.

More specific to our \ac{ddos} problem, Ghanbari et al. propose a feature extraction algorithm based on the \textit{discrete wavelet transform} and on the \textit{variance fractal dimension trajectory} to maximize the sensitivity of the \ac{cnn} in detecting \ac{ddos} attacks \cite{8482019}. The evaluation results show that the proposed approach recognises \ac{ddos} attacks with 87.35\% accuracy on the CAIDA \ac{ddos} attack dataset \cite{caida2007}. Although the authors state that their method allows real-time detection of \ac{ddos} attacks in a range of environments, no performance measurements are reported to support this claim.

DeepDefense \cite{7946998} combines \acp{cnn} and \acp{rnn} to translate original traffic traces into arrays that contain packet features collected within sliding time windows. The results presented demonstrate high accuracy in \ac{ddos} attack detection within the selected ISCX2012 dataset \cite{ISCXIDS2012}. However, it is not clear if these results were obtained on unseen test data, or are results from the training phase. Furthermore, the number of trainable parameters in the model is extremely large indicating a long and resource-intensive training phase. This would significantly challenge implementation in an online system with constrained resources, as will be discussed in Sec. \ref{sec:evaluation} and \ref{sec:usecase}.

Although deep learning offers the potential for an effective \ac{ddos} detection method, as described, existing approaches are limited by their suitability for online implementation in resource-constrained environments. In Sec. \ref{sec:evaluation}, we compare our proposed solution, \ourtool, with the state-of-the-art, specifically \cite{7946998,tr-ids,8343104,8599821,8666588} and demonstrate the contributions of \ourtool.

\section{Methodology}\label{sec:methodology}

In this paper we present \ourtool, a \ac{cnn}-based solution for \ac{ddos} detection that can be deployed in online resource-constrained environments.  Our \ac{cnn} encapsulates the learning of malicious activity from traffic to enable the identification of \ac{ddos} patterns regardless of their temporal positioning. This is a fundamental benefit of \acp{cnn}; to produce the same output regardless of where a pattern appears in the input. This encapsulation and learning of features whilst training the model removes the need for excessive feature engineering, ranking and selection.  To support an online attack detection system, we use a novel preprocessing method for the network traffic that generates a spatial data representation used as input to the \ac{cnn}. In this section, we introduce the network traffic preprocessing method, the \ac{cnn} model architecture, and the learning procedure.

\subsection{Network Traffic preprocessing}\label{sec:pcap-processing}
Network traffic is comprised of data flows between endpoints. Due to the shared nature of the communication link, packets from different data flows are multiplexed resulting in packets from the same flow being separated for transmission. This means that the processing for live presentation of traffic to a \ac{nids} is quite different to the processing of a static dataset comprising complete flows. For the same reason, the ability to generate flow-level statistics, as relied upon by many of the existing works described in Sec. \ref{sec:related}, is not feasible in an online system.

\begin{table}[!t]
	\centering
	\scriptsize
	\caption{Glossary of symbols.}
	\renewcommand{\arraystretch}{1.1}
	\label{tab:notations}
	\begin{adjustbox}{width=1\linewidth}
		\begin{tabular}{|l|l|l|l|}
			\hline
			\textit{$\alpha$} & Learning rate & \textit{n} & Number of packets per sample \\
			\hline
			\textit{f} & Number of features per packet & \textit{s} & Batch size \\
			\hline
			\textit{h} & Height of convolutional filters & \textit{t} & Time window duration \\
			\hline
			\textit{id} & 5-tuple flow identifier & $\tau$ & Time window start time\\
			\hline
			\textit{k} & Number of convolutional filters & $\mathcal{E}$ & Array of labelled samples \\
			\hline
			\textit{m} & Max pooling size & $\mathcal{L}$ & Set of labels \\
			\hline
		\end{tabular}
	\end{adjustbox}
\end{table}

In order to develop our online \ac{nids}, we created a tool that converts the traffic flows extracted from network traffic traces of a dataset into array-like data structures and splits them into sub-flows based on time windows. Shaping the input as packet flows in this manner creates a spatial data representation, which allows the \ac{cnn} to learn the characteristics of \ac{ddos} attacks and benign traffic through the convolutional filters sliding over such input to identify salient patterns. This form of input is compatible with traffic captured in online deployments. The process is illustrated in Algorithm \ref{lst:pcap-algorithm} and described next. The symbols are defined in Table \ref{tab:notations}. 

\begin{algorithm}[h!]
	\caption{Network traffic preprocessing algorithm}
	\label{lst:pcap-algorithm}
	\begin{algorithmic}[1]

		\renewcommand{\algorithmicrequire}{\textbf{Input:}}
		\renewcommand{\algorithmicensure}{\textbf{Output:}}
		\Require Network traffic trace $(NTT)$, flow-level labels $(\mathcal{L})$, time window $(t)$, max packets/sample $(n)$
		\Ensure List of labelled samples $(\mathcal{E})$
		\Procedure{PreProcessing}{$NTT$, $\mathcal{L}$, $t$, $n$}
		\State $\mathcal{E}\gets\emptyset$ \Comment Initialise the set of samples
		\State $\tau \gets -1$ \Comment Initialise the time window start-time 
		\ForAll {$pkt \in NTT $} \Comment Loop over the packets
		\State $id\gets pkt.tuple$ \Comment 5-tuple flow identifier
		\If {$\tau == -1$ \textbf{or} $pkt.time > \tau+t$} 
		\State $\tau\gets pkt.time$ \Comment Time window start time
		\EndIf
		\If {$\big|\mathcal{E}[\tau,id]\big|<n$} \Comment Max $n$ pkts/sample
		\State $\mathcal{E}[\tau,id].pkts.append(pkt.features)$
		\EndIf
		\EndFor
		\State $\mathcal{E}\gets normalization\_padding(\mathcal{E})$
		\ForAll {$e \in \mathcal{E} $} \Comment Labelling
		\State $e.label\gets \mathcal{L}[e.id]$ \Comment Apply the label to the sample
		\EndFor
		\State \textbf{return} $\mathcal{E}$
		\EndProcedure

	\end{algorithmic}
\end{algorithm}

\textbf{Feature extraction.}
Given a traffic trace file from the dataset and a pre-defined time window of length $t$ seconds, the algorithm collects all the packets from the file with capture time between $t_0$, the capture time of the first packet, and time $t_0+t$. From each packet, the algorithm extracts \features\ attributes (see Table \ref{tab:features}). We intuitively exclude those attributes that would be detrimental to the generalization of the model, such as IP addresses and TCP/UDP ports (specific to the end-hosts and user applications), link layer encapsulation type (linked to the network interfaces) and application-layer attributes (e.g., IRC or HTTP protocol attributes).

\textbf{Data processing algorithm.}
This procedure, described in Algorithm \ref{lst:pcap-algorithm} at lines 4-12, simulates the traffic capturing process of online \acp{ids}, where the traffic is collected for a certain amount of time $t$ before being sent to the anomaly detection algorithms. Hence, such algorithms must base their decisions on portions of traffic flows, without the knowledge of their whole life. To simulate this process, the attributes of the packets belonging to the same bi-directional traffic flow are grouped in chronological order to form an example of shape $[n,f]$ (as shown in Table \ref{tab:features}), where $f$ is the number of features (\features) and $n$ is the maximum number of packets the parsing process collects for each flow within the time window. $t$ and $n$ are hyper-parameters for our \ac{cnn}. Flows longer than $n$ are truncated, while shorter flows are zero-padded at the end during the next stage after normalization. The same operations are repeated for the packets within time window $[t_0+t,t_0+2t]$ and so on, until the end of the file. 

Logically, we hypothesize that short time windows enable the online systems to detect \ac{ddos} attacks within a very short time frame. Conversely, higher values of $t$ and $n$ offer more information on flows to the detection algorithms, which we expect to result in higher detection accuracy. The sensitivity of our \ac{cnn} to the values of $t$ and $n$ is evaluated in Sec. \ref{sec:setup}.

\begin{figure}[!t]
	\begin{center}
		\includegraphics[width=1\linewidth]{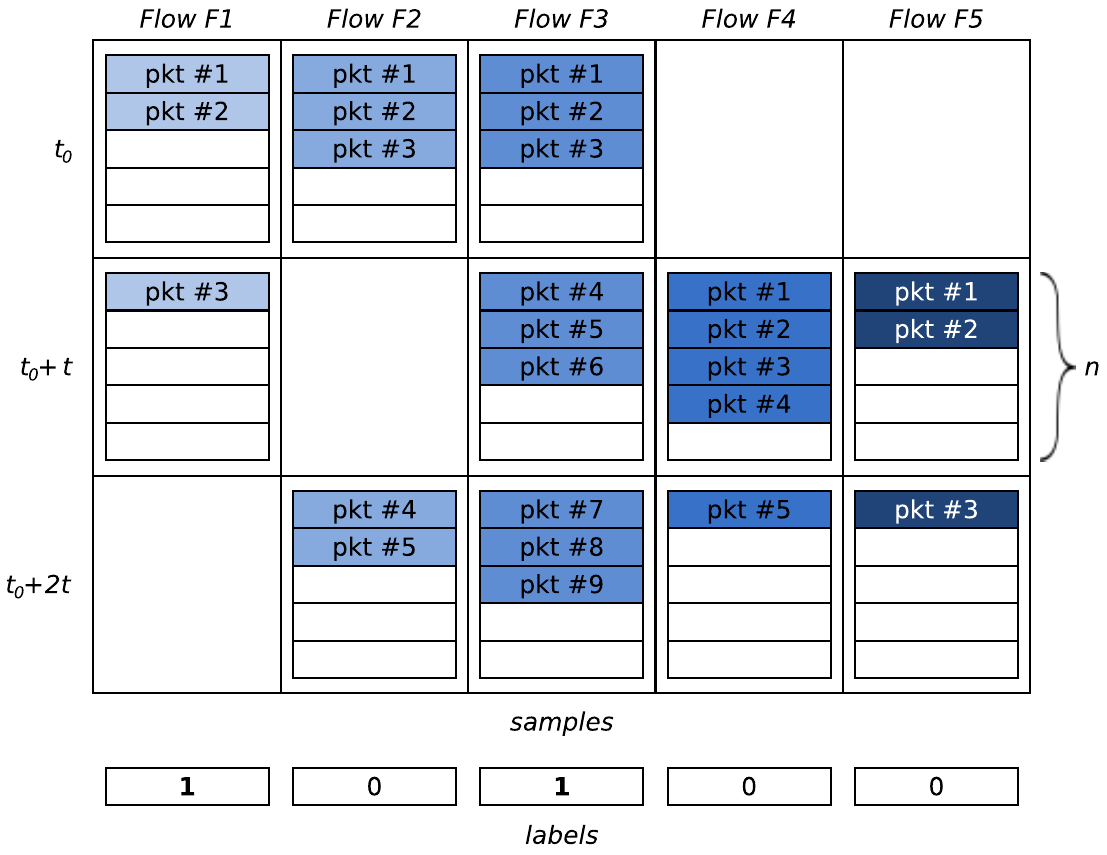}
		\caption{Graphical representation of $\mathcal{E}$.}
		\label{fig:dataset-format}
	\end{center}
\end{figure}

The output of this process can be seen as a bi-dimensional array of samples ($\mathcal{E}[\tau,id]$ in Algorithm \ref{lst:pcap-algorithm}). A row of the array represents the samples whose packets have been captured in the same time window, whilst a column represents the samples whose packets belong to the same bi-directional flow. A graphical representation of array $\mathcal{E}$ is provided in Fig. \ref{fig:dataset-format}.

\begin{table*}[!t]
	\centering
	\small
	\caption{A TCP flow sample before normalization.}
	\label{tab:features}
	\renewcommand{\arraystretch}{1.1}
	\begin{threeparttable}
		\begin{tabular}{rc|c|c|c|c|c|c|c|c|c|c|c}
			\Cline{0.8pt}{2-13}
			&
			\textbf{Pkt \#} & \begin{tabular}{@{}c@{}}\textbf{Time}\\\textbf{(sec)}\tnote{1}\end{tabular} & \begin{tabular}{@{}c@{}}\textbf{Packet}\\\textbf{Len}\end{tabular} & \begin{tabular}{@{}c@{}}\textbf{Highest}\\\textbf{Layer}\tnote{2}\end{tabular} &
			\begin{tabular}{@{}c@{}}\textbf{IP}\\\textbf{Flags}\end{tabular} &  \textbf{Protocols}\tnote{3} & \begin{tabular}{@{}c@{}}\textbf{TCP}\\\textbf{Len}\end{tabular} & \begin{tabular}{@{}c@{}}\textbf{TCP}\\\textbf{Ack}\end{tabular} & \begin{tabular}{@{}c@{}}\textbf{TCP}\\\textbf{Flags}\end{tabular} & \begin{tabular}{@{}c@{}}\textbf{TCP}\\\textbf{Window Size}\end{tabular} & \begin{tabular}{@{}c@{}}\textbf{UDP}\\\textbf{Len}\end{tabular} & \begin{tabular}{@{}c@{}}\textbf{ICMP}\\\textbf{Type}\end{tabular}  \\ \Cline{0.8pt}{2-13}
			
			\multirow{4}*{\hspace{-1em}\rotatebox[origin=c]{90}{\textit{Packets}}\vspace{0.5em}$\left\{\begin{matrix}\vspace{4.5em}\end{matrix}\right.$\hspace{-1em}}
			& 0 & 0 & 151 & 99602525 & 0x4000 & 0011010001000b & 85 & 336 & 0x018 & 1444 &  0 & 0 \\ \cline{2-13}
			& 1 & 0.092 & 135 & 99602525  & 0x4000 & 0011010001000b & 69 & 453 & 0x018 & 510 & 0 & 0 \\ \cline{2-13}
			& \vdots & \vdots & \vdots & \vdots & \vdots & \vdots & \vdots & \vdots & \vdots & \vdots & \vdots & \vdots \\  \cline{2-13}
			& $j$ & 0.513 &  66 & 78354535  & 0x4000 & 0010010001000b & 0 & 405 & 0x010 & 1444 & 0 & 0 \\ \cline{2-13}
			\multirow{3}*{\hspace{-1em}\rotatebox[origin=c]{90}{\textit{Padding}}\hspace{-0.2em}\vspace{0.5em}$\left\{\begin{matrix}\vspace{3.3em}\end{matrix}\right.$\hspace{-1em}} 
			& $j+1$ & 0 & 0 & 0 & 0 & 0000000000000b & 0 & 0 & 0 & 0 & 0 & 0 \\ \cline{2-13}
			& \vdots & \vdots & \vdots & \vdots & \vdots & \vdots & \vdots & \vdots & \vdots & \vdots & \vdots & \vdots \\  \cline{2-13}
			& $n$ & 0 & 0 & 0 & 0 & 0000000000000b & 0 & 0 & 0 & 0 & 0 & 0 \\ 
			\Cline{0.8pt}{2-13}
		\end{tabular}
		\vspace{0.2em}
		\begin{tablenotes}
			\scriptsize
			\setlength{\itemindent}{0.5em}
			\item[1] Relative time from the first packet of the flow.
			\item[2] Numerical representation of the highest layer recognised in the packet.
			\item[3] Binary representation of the list of protocols recognised in the packet using the well-known \ac{bow} model. It includes protocols from Layer 2 (\textit{arp})
			\item[] to common clear text application layer protocols such as \textit{http}, \textit{telnet}, \textit{ftp} and \textit{dns}. 
		\end{tablenotes}
	\end{threeparttable}	
\end{table*}

\textbf{Normalization and padding.}
Each attribute value is normalized to a $[0,1]$ scale and the samples are zero-padded so that each sample is of fixed length $n$, since having samples of fixed length is a requirement for a CNN to be able to learn over a full sample set.  In Fig. \ref{fig:dataset-format}, each non-empty element of the array $\mathcal{E}$ is a compact graphical representation of a sample. 
In each $\mathcal{E}$ element, coloured rows are the packets in the form of \features\ normalized attributes (i.e., the upper part of Table \ref{tab:features}), while the white rows represent the zero-padding (i.e., the lower part of Table \ref{tab:features}). Please note that, empty elements in Fig. \ref{fig:dataset-format} are for visualization only and are not included in the dataset. An empty $\mathcal{E}[\tau,id]$ means that no packets of flow $id$ have been captured in time window $[\tau,\tau+t]$ (e.g. $\mathcal{E}[t_0,F4]$).

\textbf{Labelling.} Each example $\mathcal{E}[\tau,id]$ is labelled by matching its flow identifier $id$ with the labels provided with the original dataset (lines 14-16 in Algorithm \ref{lst:pcap-algorithm}). This also means that the value of the label is constant along each column of array $\mathcal{E}$, as represented in Fig. \ref{fig:dataset-format}.

\subsection{\ourtool\ Model Architecture}\label{sec:architecture}
\begin{figure}[]
	\begin{center}
		\includegraphics[width=1\linewidth]{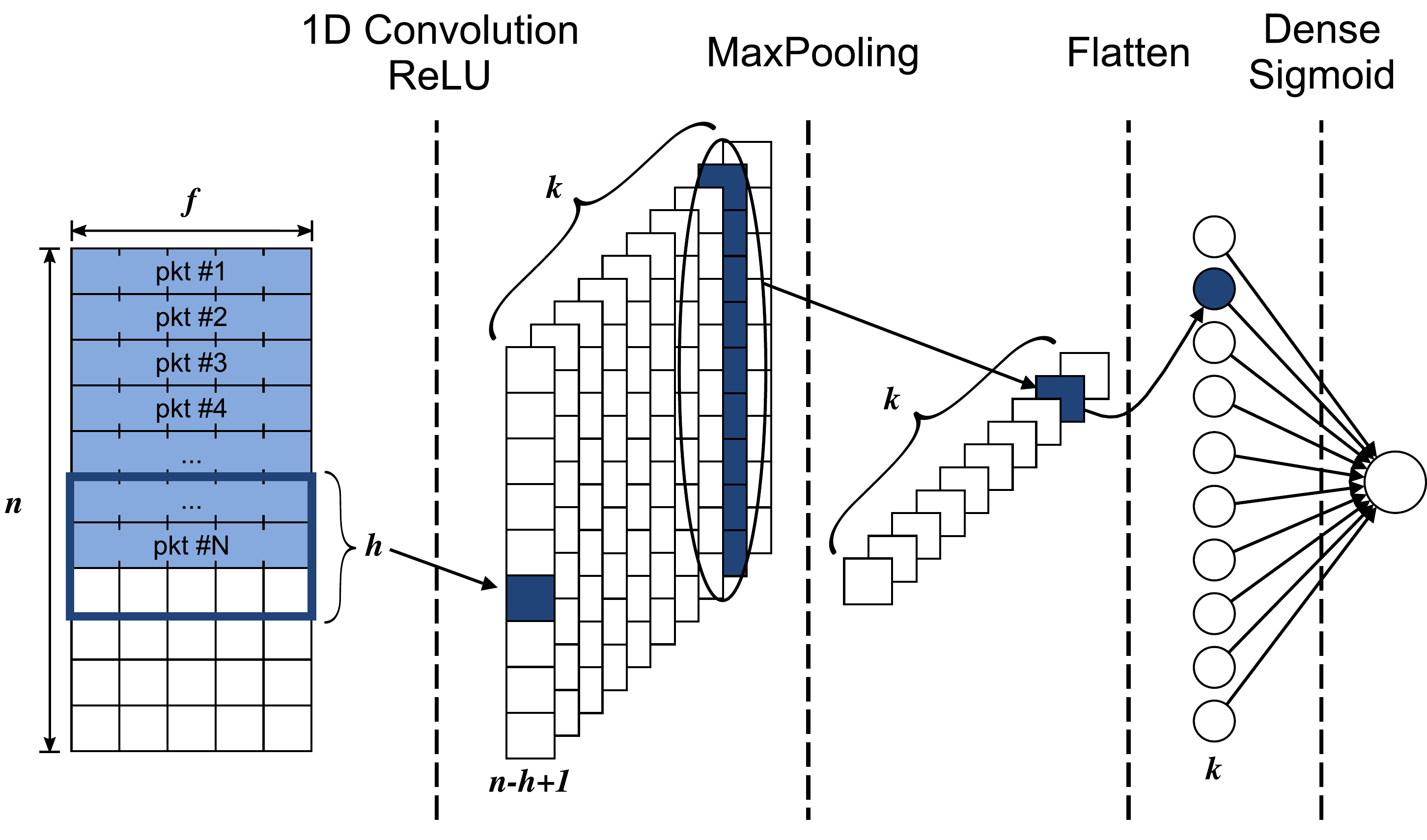}
		\caption{\ourtool\ architecture.}
		\label{fig:CNN-architecture}
	\end{center}
\end{figure}

We take the output from Algorithm \ref{lst:pcap-algorithm} as input to our \ac{cnn} model for the purposes of online attack detection.  \ourtool\ classifies traffic flows into one of two classes, either malicious (\ac{ddos}) or benign.  Our objective is to minimise the complexity and performance time of this \ac{cnn} model for feasible deployment on resource-constrained devices.  To achieve this, the proposed approach is a lightweight, supervised detection system that incorporates a \ac{cnn}, similar to that of \cite{KimConvolutionalNN} from the field of Natural Language Processing.  CNNs have shared and reused parameters with regard to the weights of the kernels, whereas in a traditional neural network every weight is used only once. This reduces the storage and memory requirements of our model.  The complete architecture is depicted in Fig. \ref{fig:CNN-architecture} and described in the next sections, with the hyper-parameter tuning and ablation studies being discussed in Sec. \ref{sec:setup}.

\textbf{Input layer.}
Recall that each traffic flow has been reshaped into a 2-D matrix of packet features as per Sec. \ref{sec:pcap-processing}, creating a novel spatial representation that enables the \ac{cnn} to learn the correlation between packets of the same flow.  Thus, this first layer takes as input a traffic flow represented by a matrix $F$ of size $n\times f$.  $F$ contains $n$ individual packet vectors, such that $F$ = \{$pkt_1$, ... , $pkt_n$\} where $pkt_n$ is the $n$th packet in a flow, and each packet vector has length $f$ = \features\ features.

\textbf{CNN layer.} As per Fig. \ref{fig:CNN-architecture}, each input matrix $F$ is operated on by a single convolutional layer with $k$ filters of size $h\times f$, with $h$ being the length of each filter, and again $f$ = \features.  Each filter, also known as a kernel or sliding window, convolves over $F$ with a step of 1 to extract and learn local features that contain useful information for detection of \ac{ddos} and benign flows.  Each of the $k$ filters generates an activation map $a$ of size ($n-h+1$), such that $a_k = ReLU(Conv(F)W_k,b_k)$, where $W_k$ and $b_k$ are the weight and bias parameters of the $k$th filter that are learned during the training stage.  To introduce non-linearity among the learned filters, we use the rectified linear activation function $ReLU(x) = max\{0,x\}$, as per convention for CNNs.  All activation maps are stacked, creating an activation matrix $A$ of size $(n-h+1)\times k$, such that $A$ = $[a_1 | ... | a_k]$.  

There are two main benefits of including a \ac{cnn} in our architecture.  Firstly, it allows the model to benefit from efficiency gains compared to standard neural networks, since the weights in each filter are reused across the whole input. Sharing weights, instead of the full end-to-end connectivity with a standard neural net, makes the model more lightweight and reduces its memory footprint as the number of learnable parameters is greatly reduced.  Secondly, during the training phase, the \ac{cnn} automatically learns the weights and biases of each filter such that the learning of salient characteristics and features is encapsulated inside the resulting model during training. This reduces the time-consuming feature engineering and ranking involved in statistical and traditional machine learning methods, which relies on expert human knowledge. As a result, this model is more adaptable to new subtleties of \ac{ddos} attack, since the training stage can be simply repeated anytime with fresh training data without having to craft and rank new features.

\textbf{Max pooling layer.}
For max pooling, we down-sample along the first dimension of $A$, which represents the temporal nature of the input.  A pool size of $m$ produces an output matrix $m_o$ of size $((n-h+1)/m)\times k$, which contains the largest $m$ activations of each learned filter, such that $m_o = [max(a_1) | ... | max(a_k)]$.  In this way, the model disregards the less useful information that produced smaller activations, instead paying attention to the larger activations.  This also means that we dispose of the positional information of the activation, i.e. where it occurred in the original flow, giving a more compressed feature encoding, and, in turn, reducing the complexity of the network.  $m_o$ is then flattened to produce the final one-dimensional feature vector $v$ to be input to the classification layer.

\textbf{Classification layer.}
$v$ is input to a fully-connected layer of the same size, and the output layer has a sole node.  This output $x$ is passed to the sigmoid activation function such that $\sigma(x) = 1/(1+e^{-x})$.  This constrains the activation to a value of between 0 and 1, hence returning the probability $p\in[0,1]$ of a given flow being a malicious \ac{ddos} attack. The flow is classified as \ac{ddos} when $p>0.5$, and benign otherwise.

\subsection{The Learning Procedure}\label{sec:learning}
When training \ourtool{}, the objective is to minimise its cost function through iteratively updating all the weights and biases contained within the model.  These weights and biases are also known as trainable, or learnable, parameters.  The cost function calculates the cost, also called the error or the loss, between the model's prediction, and the ground truth of the input.  Hence by minimising this cost function, we reduce the prediction error.  At each iteration in training, the input data is fed forward through the network, the error calculated, and then this error is back-propagated through the network. This continues until convergence is reached, when further updates don't reduce the error any further, or the training process reaches the set maximum number of epochs.  With two classes in our problem the binary cross-entropy cost function is used.  Formally this cost function $c$ that calculates the error over a batch of $s$ samples can be written as: 
\begin{equation}
c = -\frac{1}{s}\sum_{j=1}^{s}(y_j\log p_j + (1-y_j)\log(1-p_j)) 
\end{equation}

where $y_j$ is the ground truth target label for each flow $j$ in the batch of $s$ samples, and $p_j$ is the predicted probability flow $j$ is malicious \ac{ddos}.  This is supervised learning because each flow in our datasets is labelled with the ground truth, either \ac{ddos} or benign.  To reduce bias in our learning procedure, we ensure that these datasets are balanced with equal numbers of malicious and benign flows, which gives a greater degree of confidence that the model is learning the correct feature representations from the patterns in the traffic flows.  As previously highlighted, the learning is encapsulated inside the model by all the weights and biases, meaning that our approach does not require significant expert input to craft bespoke features and statistically assess their importance during preprocessing, unlike many existing methods, as outlined in Sec. \ref{sec:related}. 
\section{Experimental Setup}\label{sec:setup}
\pgfplotstableread[header=true]{
P	T=1		T=10	T=100
1	0.94244	0.91464	0.88712
2	0.97988	0.98657	0.98975
3	0.98695	0.99104	0.9942
4	0.99045	0.99051	0.99379
5	0.9912	0.99202	0.99367
10	0.99022	0.99172	0.99406
20	0.98998	0.9923	0.99393
50	0.99023	0.99451	0.99472
100	0.98844	0.99153	0.99501

}\FScoreP

\pgfplotstableread[header=true]{
T	P=1		P=2		P=10	P=100
1	0.94244	0.98975	0.99022	0.98844
2	0.92951	0.98569	0.99179	0.98957
3	0.93132	0.98486	0.99202	0.98887
4	0.93276	0.98657	0.98756	0.98911
5	0.9272	0.97988	0.98678	0.98967
10	0.91464	0.98448	0.99172	0.99153
20	0.90821	0.98295	0.99325	0.9937
50	0.89324	0.98202	0.9946	0.99383
100	0.88712	0.98531	0.99406	0.99501

}\FScoreT

\pgfplotstableread[header=true]{
	BATCHSIZE	PKTSECGPU	SAMPLESECGPU	PKTSECCPU	SAMPLESECCPU
	64	504809	18780	346298	12883
	128	458845	16710	396203	14740
	256	562915	20942	445924	16589
	512	591070	21989	479257	17829
	1024	614937	22877	521698	19408
	2048	587934	21872	496299	18463
	4096	507707	18888	448861	16699
	8192	501544	18658	416435	15492

}\PredictionTX

\subsection{Datasets}\label{sec:datasets}
Our \ac{cnn} model is validated with recent datasets ISCX2012 \cite{ISCXIDS2012}, CIC2017 \cite{CICIDS2017} and CSECIC2018 \cite{CSECICIDS2018} provided by the Canadian Institute for Cybersecurity of the \ac{unb}, Canada. They consist of several days of network activity, normal and malicious, including \ac{ddos} attacks. The three datasets are publicly available in the form of traffic traces in \textit{pcap} format including full packet payloads, plus supplementary text files containing the labels and statistical details for each traffic flow. 

The \ac{unb} researchers have generated these datasets by using profiles to accurately represent the abstract properties of human and attack behaviours. One profile characterises the normal network activities and provides distribution models for applications and  protocols (HTTP, SMTP, SSH, IMAP, POP3, and FTP) produced with the analysis of real traffic traces. Other profiles describe a variety of attack scenarios based on recent security reports. They are used to mimic the behaviour of the malicious attackers by means of custom botnets and well-known \ac{ddos} attacking tools such as High Orbit Ion Cannon (HOIC)~\cite{hoic} and its predecessor, the Low Orbit Ion Cannon (LOIC)~\cite{loic}. HOIC and LOIC have been widely used by Anonymous and other hacker groups in some highly-publicized attacks against PayPal, Mastercard, Visa, Amazon, Megaupload, among others \cite{paypal}. 

Table \ref{tab:unb-datasets} shows the parts of the three datasets used in this work. In the table, the column \textit{Traffic trace} specifies the name of the trace, according to \cite{ISCXIDS2012}, \cite{CICIDS2017} and \cite{CSECICIDS2018}. Specifically, the \textit{ISCX2012-Tue15} trace contains a DDoS attack based on an IRC botnet. The \textit{CIC2017-Fri7PM} trace contains a HTTP \ac{ddos} generated with LOIC, while the \textit{CSECIC2018-Wed21} trace contains a HTTP \ac{ddos} generated with HOIC. With respect to the original file, the trace \textit{CIC2017-Fri7PM} is reduced to timeslot 3.30PM-5.00PM to exclude malicious packets related to other cyber attacks (port scans and backdoors). 

\begin{table}[h!]
	\caption{The datasets from \ac{unb} \cite{unb-datasets}.} 
	\label{tab:unb-datasets}
	\small 
	\centering 
	\begin{tabular}{llccc} \toprule[\heavyrulewidth]
		\textbf{Dataset} & \textbf{Traffic trace} & \textbf{\#Flows} & \textbf{\#Benign} & \textbf{\#\ac{ddos}} \\ \midrule[\heavyrulewidth]
		ISCX2012	& Tue15 & 571698 & 534320 & 37378\\ \midrule
		CIC2017	& Fri7PM & 225745 & 97718 & 128027 \\ \midrule
		CSECIC2018& Wed21 & 1048575 & 360832 & 687743\\ \bottomrule[\heavyrulewidth]
	\end{tabular}
\end{table}

In an initial design, the model was trained and validated on the ISCX2012 dataset producing high accuracy results. However, testing the model on the CIC2017 dataset confirmed the generally held observation that a model trained on one dataset will not necessarily perform well on a completely new dataset. In particular, we obtained a false negative rate of about 17\%. This can be attributed to the different attacks represented in the two datasets, as previously described. What we attempt in this work is to develop a model that when trained and validated across a mixed dataset can reproduce the high performance results on completely unseen test data. To achieve this, a combined training dataset is generated as described in Sec. \ref{sec:data-preparation}.

\subsection{Data preparation}\label{sec:data-preparation}
We extract the 37378 \ac{ddos} flows from ISCX2012, plus randomly select 37378 benign flows from the same year to balance. We repeat this process with 97718/97718 benign/\ac{ddos} flows for CIC2017 and again with 360832/360832 benign/\ac{ddos} flows for CSECIC2018. 

After the pre-preprocessing stage, where flows are translated into array-like data structures (Sec. \ref{sec:pcap-processing}), each of the three datasets is split into training (90\%) and test (10\%) sets, with 10\% of the training set used for validation. Please note that, the split operation is performed on a per-flow basis to ensure that samples obtained from the same traffic flow end up in the same split, hence avoiding the ``contamination'' of the validation and test splits with data used for the training. We finally combine the training splits from each year by balancing them with equal proportions from each year to produce a single training set. We do the same with the validation and test splits, to obtain a final dataset referred to as \textit{UNB201X} in the rest of the paper. UNB201X training and validation sets are only used for training the model and tuning the hyper-parameters (Sec. \ref{sec:tuning}), while the test set is used for the evaluation presented in Sec. \ref{sec:evaluation} and \ref{sec:usecase}, either as a whole combined test set, or as individual per-year test sets for state-of-the-art comparison.  

A summary of the final UNB201X splits is presented in Table \ref{tab:datasets-splits}, which reports the number of samples as a function of time window duration $t$. As illustrated in Table \ref{tab:datasets-splits}, low values of this hyper-parameter yield larger numbers of samples. Intuitively, using short time windows leads to splitting traffic flows into many small fragments (ultimately converted into samples), while long time windows produce the opposite result. In contrast, the value of $n$ has a negligible impact on the final number of samples in the dataset.

\begin{table}[h!]
	\caption{UNB201X dataset splits.} 
	\label{tab:datasets-splits}
	\small 
	\centering 
	\begin{tabular}{lcccc} \toprule[\heavyrulewidth]
		\begin{tabular}{@{}l@{}}\textbf{Time}\\\textbf{Window}\end{tabular} & \begin{tabular}{@{}c@{}}\textbf{Total}\\\textbf{Samples}\end{tabular} & \begin{tabular}{@{}c@{}}\textbf{Training}\end{tabular} & \begin{tabular}{@{}c@{}}\textbf{Validation}\end{tabular}& \begin{tabular}{@{}c@{}}\textbf{Test}\end{tabular} \\ \midrule[\heavyrulewidth]
	
		$t$=1s & 480519 & 389190 & 43272 & 48057\\   
		$t$=2s & 353058 & 285963 & 31782 & 35313\\ 
		$t$=3s & 310590 & 251574 & 27957 & 31059\\ 
		
		$t$=4s & 289437 & 234438 & 26055 & 28944 \\   
		$t$=5s & 276024 & 223569 & 24852 & 27603\\ 
		$t$=10s & 265902 & 215379 & 23931 & 26592\\ 
	
		$t$=20s & 235593 & 190827 & 21204 & 23562\\   
		$t$=50s & 227214 & 184041 & 20451 & 22722\\ 
		$t$=100s & 224154 & 181551 & 20187 & 22416\\ \bottomrule[\heavyrulewidth]
	\end{tabular}
\end{table}

\subsection{Evaluation methodology}
As per convention in the literature, we report the metrics \textit{Accuracy (ACC)}, \textit{\ac{fpr}}, \textit{Precision} (or \textit{\ac{ppv}}), \textit{Recall} (or \textit{\ac{tpr}}) and \textit{F1 Score (F1)}, with a focus on the latter. \textit{Accuracy} is the percentage of correctly classified samples (both benign and \ac{ddos}). \textit{\ac{fpr}} represents the percentage of samples that are falsely classified as \ac{ddos}. \textit{\ac{ppv}} is the ratio between the correctly detected \ac{ddos} samples and all the detected \ac{ddos} samples (true and false). \textit{\ac{tpr}} represents the percentage of \ac{ddos} samples that are correctly classified as such. The \textit{F1 Score} is an overall measure of a model's performance; that is the harmonic mean of the \textit{\ac{ppv}} and \textit{\ac{tpr}}. These metrics are formally defined as follows:

\begin{center}
\scalebox{1}{$ACC=\frac{TP+TN}{TP+TN+FP+FN}\ \ FPR=\frac{FP}{FP+TN}$}\\ \vspace{0.8em}
\scalebox{1}{$PPV=\frac{TP}{TP+FP}\ \ TPR=\frac{TP}{TP+FN}\ \ F1=2\cdot\frac{PPV\cdot TPR}{PPV + TPR}$}
\vspace{1mm}
\end{center}

\noindent where \textit{TP=True Positives}, \textit{TN=True Negatives}, \textit{FP=False Positives}, \textit{FN=False Negatives}.    

The output of the training process is a combination of trainable and hyper parameters that maximizes the \textit{F1 Score} on the validation set or, in other words, that minimizes the total number of False Positives and False Negatives. 

Model training and validation have been performed on a server-class computer equipped with two 16-core Intel Xeon Silver 4110 @2.1\,GHz CPUs and 64\,GB of RAM. The models have been implemented in Python v3.6 using the Keras API v2.2.4 \cite{keras} on top of Tensorflow 1.13.1 \cite{tensorflow}.

\subsection{Hyper-parameter tuning}\label{sec:tuning}
Tuning the hyper-parameters is an important step to optimise the model's accuracy, as their values influence the model complexity and the learning process.  Prior to our experiments, we empirically chose the hyper-parameter values based on the results of preliminary tuning and on the motivations described per parameter. We then adopted a grid search strategy to explore the set of hyper-parameters using \textit{F1 score} as the performance metric.  At each point in the grid, the training continues indefinitely and stops when the loss does not decrease for a consecutive 25 times. Then, the search process saves the \textit{F1 score} and moves to the next point.

As per Sec. \ref{sec:data-preparation}, UNB201X is split into training, validation and testing sets.  For hyper-parameter tuning, we use only the validation set. It is important to highlight that we do not tune to the test set, as that may artificially improve performance. 
The test set is kept completely unseen, solely for use in generating our experimental results, which are reported in Sec. \ref{sec:evaluation}.

\textbf{Maximum number of packets/sample.}
$n$ is important for the characterization of the traffic and for capturing the temporal patterns of traffic flows. The value of $n$ indicates the maximum number of packets of a flow recorded in chronological order in a sample. 

\begin{figure}[h!]
	\begin{tikzpicture}
	\begin{semilogxaxis}[  
	legend pos=south east,
	legend columns=1,
	height=5 cm,
	width=1\linewidth,
	grid = both,
	xlabel={Value of hyper-parameter $n$ (packets/example) in logarithmic scale},
	ylabel={F1 Score},
	scaled y ticks=false,
	scaled x ticks=false,
	xmin=1,
	xmax=100,
	xtick={1,2,3,4,5,10,20,50,100},
	xticklabels={1,2,3,4,5,10,20,50,100},
	xtick pos=left,
	ymin=0.88, ymax=1,
	ytick={0.88,0.90,0.95,0.98,0.99,1},
	yticklabels={0.88,0.90,0.95,0.98,0.99,1},
	ytick pos=left,
	enlargelimits=0.02,
	]
	\addplot [color=red,style=semithick,mark=*,mark size=1.4] table[x index=0,y index=1] {\FScoreP};
	\addplot [color=blue,style=semithick,mark=x,mark size=2] table[x index=0,y index=2] {\FScoreP};
	\addplot [color=orange,style=semithick,mark=triangle*,mark size=1.8] table[x index=0,y index=3] {\FScoreP};
	\node[fill=white,anchor=west] at (axis cs: 1.8,.96) {\sffamily\scriptsize $\alpha=0.01$, $s=2048$, $k=64$, $h=3$, $m=n-h+1$};
	\legend{$t$=1, $t$=10,$t$=100}
	\end{semilogxaxis}
	\end{tikzpicture}
	\caption{Sensitivity of our model to hyper-parameter $n$.}
	\label{fig:sensitivity_n}
\end{figure}
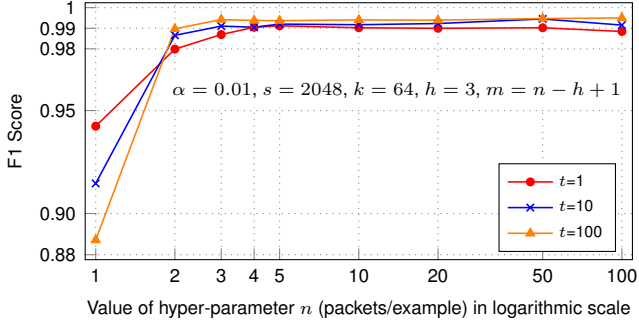

The resulting set of packets describes a portion of the life of the flow in a given time window, including the (relative) time information of packets. Repetition-based \ac{ddos} attacks use a small set of messages at approximately constant rates, therefore a small value of $n$ is sufficient to spot the temporal patterns among the packet features, hence requiring a limited number of trainable parameters. On the other hand, more complex attacks, such as the ones performed with the HOIC tool, which uses multiple HTTP headers to make the requests appear legitimate, might require a larger number of packets to achieve the desired degree of accuracy. Given the variety of \ac{ddos} tools used to simulate the attack traffic in the dataset (IRC-based bot, LOIC and HOIC), we experimented with $n$ ranging between 1 and 100, and we compared the performance in terms of \textit{F1 score}. The results are provided in Fig. \ref{fig:sensitivity_n} for different durations of time window $t$, but at fixed values of the other hyper-parameters for the sake of visualisation. 

The \textit{F1 score} steadily increases with the value of $n$ when $n<5$, and then stabilises when $n\ge 5$. However, an increase in \textit{F1 score} is still observed up to $n=100$. 
Although, a low value of $n$  can be used to speed up the detection time (less convolutions) and to reduce the requirements in terms of storage and RAM (smaller sample size), which links to our objective of a lightweight implementation, we wish to balance high accuracy with low resource consumption. This will be demonstrated in Sec. \ref{sec:usecase}.

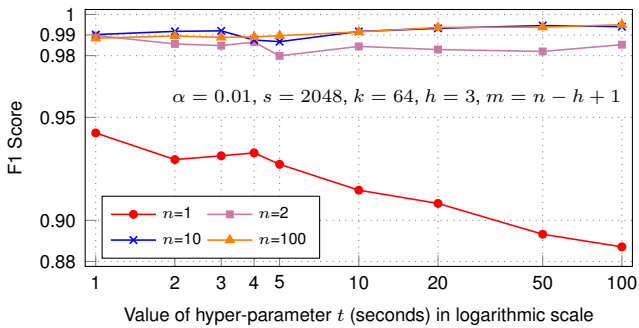
\begin{figure}[h!]
	\begin{tikzpicture}
	\begin{semilogxaxis}[  
	legend pos=south west,
	legend columns=2,
	height=5 cm,
	width=1\linewidth,
	grid = both,
	xlabel={Value of hyper-parameter $t$ (seconds) in logarithmic scale},
	ylabel={F1 Score},
	scaled y ticks=false,
	scaled x ticks=false,
	xmin=1,
	xmax=100,
	xtick={1,2,3,4,5,10,20,50,100},
	xticklabels={1,2,3,4,5,10,20,50,100},
	xtick pos=left,
	ymin=0.88, ymax=1,
	ytick={0.88,0.90,0.95,0.98,0.99,1},
	yticklabels={0.88,0.90,0.95,0.98,0.99,1},
	ytick pos=left,
	enlargelimits=0.02,
	]
	\addplot [color=red,style=semithick,mark=*,mark size=1.4] table[x index=0,y index=1] {\FScoreT};
	\addplot [color=reddish-purple,style=semithick,mark=square*,mark size=1.2] table[x index=0,y index=2] {\FScoreT};
	\addplot [color=blue,style=semithick,mark=x,mark size=2] table[x index=0,y index=3] {\FScoreT};
	\addplot [color=orange,style=semithick,mark=triangle*,mark size=1.8] table[x index=0,y index=4] {\FScoreT};
	\node[fill=white,anchor=west] at (axis cs: 1.8,.96) {\sffamily\scriptsize $\alpha=0.01$, $s=2048$, $k=64$, $h=3$, $m=n-h+1$};
	\legend{$n$=1,$n$=2,$n$=10,$n$=100}
	\end{semilogxaxis}
	\end{tikzpicture}
	\caption{Sensitivity of our model to hyper-parameter $t$.}
	\label{fig:sensitivity_t}
\end{figure}

\textbf{Time Window.}
The time window $t$ is used to simulate the capturing process of online systems (see Sec. \ref{sec:pcap-processing}). We evaluated the \textit{F1 score} for time windows ranging between 1 and 100 seconds (as in the related work e.g. \cite{7946998}) at different values of $n$. The results are shown in Fig. \ref{fig:sensitivity_t}. 

Although the number of samples in the training set decreases when $t$ increases (see Table \ref{tab:datasets-splits}), the \ac{cnn} is relatively insensitive to this hyper-parameter for $n>1$. With $n=1$, the traffic flows are represented by samples of shape $[1,f]$, i.e. only one packet/sample, irrespective of the duration of the time window. In such a corner case, since the \ac{cnn} cannot correlate the attributes of different packets within the same sample, the \textit{F1 score} is more influenced by the number of samples in the training set (the more samples, the better).   
 
\textbf{Height of convolutional filters.}
$h$ determines the height of the filters (the width is fixed to \features, the number of features), i.e. the number of packets to involve in each matrix operation.  
Testing with $h=1,2,3,4,5$, we observed a small, but noticeable, difference in the \textit{F1 score} between $h=1$ (0.9934) and $h=3$ (0.9950), with no major improvement beyond $h=3$.

\textbf{Number of convolutional filters.}
As per common practice, we experimented by increasing the number of convolutional filters $k$ by powers of 2, from $k=1$ to $k=64$. We observed a steady increase in the \textit{F1 score} with the value of $k$, which is a direct consequence of the increasing number of trainable parameters in the model.

\textbf{Resulting hyper-parameter set.}
After conducting a comprehensive grid search on 2835 combinations of hyper-parameters, we have selected the \ac{cnn} model configuration that maximises the \textit{F1 score} on the UNB201X validation set (Table \ref{tab:training_results}). That is: 
$$\textbf{n}=100,\ \textbf{t}=100,\ \textbf{k}=64,\ \textbf{h}=3,\ \textbf{m}=98$$
The resulting model, trained with batch size $s=2048$ and using the Adam optimizer \cite{adam} with learning rate $\alpha=0.01$, consists of 2241 trainable parameters, 2176 for the convolutional layer ($h\cdot f$ units for each filter plus bias, multiplied by the number of filters K) and 65 for the fully connected layer (64 units plus bias).  

As previously noted, other configurations may present lower resource requirements at the cost of a minimal decrease in \textit{F1 score}. For example, using $k=32$ would reduce the number of convolutions by half, while $n=10,20,50$ would also require fewer convolutions and a smaller memory footprint. However, setting $n=100$ not only maximises the \textit{F1 score}, but also enables a fair comparison with state-of-the-art approaches such as DeepDefense \cite{7946998} (Sec. \ref{sec:evaluation}), where the authors trained their neural networks using $n=100$ (in \cite{7946998}, the hyper-parameter is denoted as $T$). Furthermore, the chosen configuration enables a worst-case analysis for resource-constrained scenarios such as that presented in Sec. \ref{sec:usecase}.

These hyper-parameters are kept constant throughout our experiments presented in Sec. \ref{sec:evaluation} and \ref{sec:usecase}.
\begin{table}[h!]
	\caption{Scores obtained on the UNB201X validation set.} 
	\label{tab:training_results}
	\small 
	\centering 
	\begin{threeparttable}
		\begin{tabular}{lccccc} \toprule[\heavyrulewidth]
			\textbf{Validation set}  & \textbf{ACC} & \textbf{FPR} & \textbf{PPV} & \textbf{TPR} & \textbf{F1} \\ \midrule[\heavyrulewidth]
			UNB201X & 0.9950 & 0.0083 & 0.9917 & 0.9983 & 0.9950  \\ \bottomrule[\heavyrulewidth]
		\end{tabular}
	\end{threeparttable}
\end{table}

\section{Results}\label{sec:evaluation}
\pgfplotstableread[header=true]{
P	T=1		T=10	T=100
1	0.94244	0.91464	0.88712
2	0.97988	0.98657	0.98975
3	0.98695	0.99104	0.9942
4	0.99045	0.99051	0.99379
5	0.9912	0.99202	0.99367
10	0.99022	0.99172	0.99406
20	0.98998	0.9923	0.99393
50	0.99023	0.99451	0.99472
100	0.98844	0.99153	0.99501

}\FScoreP

\pgfplotstableread[header=true]{
T	P=1		P=2		P=10	P=100
1	0.94244	0.98975	0.99022	0.98844
2	0.92951	0.98569	0.99179	0.98957
3	0.93132	0.98486	0.99202	0.98887
4	0.93276	0.98657	0.98756	0.98911
5	0.9272	0.97988	0.98678	0.98967
10	0.91464	0.98448	0.99172	0.99153
20	0.90821	0.98295	0.99325	0.9937
50	0.89324	0.98202	0.9946	0.99383
100	0.88712	0.98531	0.99406	0.99501

}\FScoreT

\pgfplotstableread[header=true]{
	BATCHSIZE	PKTSECGPU	SAMPLESECGPU	PKTSECCPU	SAMPLESECCPU
	64	504809	18780	346298	12883
	128	458845	16710	396203	14740
	256	562915	20942	445924	16589
	512	591070	21989	479257	17829
	1024	614937	22877	521698	19408
	2048	587934	21872	496299	18463
	4096	507707	18888	448861	16699
	8192	501544	18658	416435	15492

}\PredictionTX

In this section, we present a detailed evaluation of the proposed approach with the datasets presented in Sec. \ref{sec:datasets}.
Evaluation metrics of  \textit{Accuracy (ACC)}, \textit{\acf{fpr}}, \textit{Precision} (\textit{\ac{ppv}}), \textit{Recall} (\textit{\ac{tpr}}) and \textit{F1 Score (F1)} have been used for performance measurement and for comparison with state-of-the-art models. 

\subsection{Detection accuracy}\label{sec:evaluation-datasets}
In order to validate our approach and the results obtained on the validation dataset, we measure the performance of \ourtool{} in classifying unseen traffic flows as benign or malicious (\ac{ddos}). Table \ref{tab:detection_performance} summarizes the results obtained on the various test sets produced through the procedure described in Sec. \ref{sec:data-preparation}. As illustrated, the very high performance is maintained across the range of test datasets indicating the robustness of the \ourtool{} design. These results are further discussed in Sec. \ref{sec:evaluation-sota}, where we compare our solution with state-of-the-art works reported in the scientific literature.

\begin{table}[h!]
	\caption{\ourtool\ detection performance on the test sets.} 
	\label{tab:detection_performance}
	\small 
	\centering 
	\begin{threeparttable}
		\begin{tabular}{lccccc} \toprule[\heavyrulewidth]
			\textbf{Test set}  & \textbf{ACC} & \textbf{FPR} & \textbf{PPV} & \textbf{TPR} & \textbf{F1} \\ \midrule[\heavyrulewidth]
			ISCX2012 & 0.9888 & 0.0179 & 0.9827 & 0.9952 & 0.9889 \\ \midrule
			CIC2017  & 0.9967 & 0.0059 & 0.9939 & 0.9994 & 0.9966 \\ \midrule
			CSECIC2018 & 0.9987 & 0.0016 & 0.9984 & 0.9989 & 0.9987 \\ \midrule
			UNB201X & 0.9946 & 0.0087 & 0.9914 & 0.9979 & 0.9946  \\ \bottomrule[\heavyrulewidth]
		\end{tabular}
	\end{threeparttable}
\end{table}

The results show that thanks to the properties of its \ac{cnn}, \ourtool{} learns to distinguish between patterns of malicious \ac{ddos} behaviour and benign flows.  Given the properties of convolutional methods, these patterns are recognised regardless of the position they occupy in a flow, demonstrating that our spatial representation of a flow is robust.  Irrespective of whether the \ac{ddos} event appears at the start or the end of the input, \ourtool{} will produce the same representation in its output.  Although the temporal dynamics in \ac{ddos} attacks might suggest that alternative \ac{dl} architectures may seem more suitable (e.g. \ac{lstm}), our novel preprocessing method combined with the \ac{cnn} removes the requirement for the model to maintain temporal context of each whole flow as the data is pushed through the network. In comparison, \acp{lstm} are known to be very difficult to train, and their performance is inherently slower for long sequences compared to \acp{cnn}.

\subsection{State-Of-The-Art Comparison}\label{sec:evaluation-sota}
For a fair comparison between \ourtool\ and the state-of-the-art, we focus our analysis on solutions that have validated the \ac{unb} datasets for \ac{ddos} attack detection. 

We have paid particular attention to DeepDefense \cite{7946998} as, similar to our approach, the model is trained with packet attributes rather than flow-level statistics used in other works. DeepDefense  translates the \textit{pcap} files of ISCX2012 into arrays that contain packet attributes collected within sliding time windows. The label assigned to a sample is the label of the last packet in the time window, according to the labels provided with the original dataset. The proposed data preprocessing technique is similar to \ourtool's. However, in \ourtool, a sample corresponds to a single traffic flow, whereas in DeepDefense a sample represents the traffic collected in a time window. 

Of the four \ac{dl} models presented in the DeepDefense paper, the one called 3LSTM produces the highest scores in the classification of \ac{ddos} traffic. Therefore, we have implemented 3LSTM for comparison purposes. The architecture of this model includes 6 LSTM layers of 64 neurons each, 2 fully connected layers of 128 neurons each, and 4 batch normalization layers. To directly compare the \ac{dl} models, we have trained 3LSTM on the UNB201X training set with $n=100$ and $t=100$ as done with \ourtool.  We have compared our implementation of 3LSTM with \ourtool\ on each of the four test sets, and present the F1 score results in Table \ref{tab:deepdefense_f1_comparison}. 

\begin{table}[h!]
	\caption{\ourtool-DeepDefense comparison (F1 score).} 
	\label{tab:deepdefense_f1_comparison}
	\small 
	\centering 
	\begin{threeparttable}
		\begin{adjustbox}{width=0.49\textwidth}
		\begin{tabular}{lccccc} \toprule[\heavyrulewidth]
			\textbf{Model} & \begin{tabular}{@{}c@{}}\textbf{Trainable}\\\textbf{Parameters}\end{tabular} & \begin{tabular}{@{}c@{}}\textbf{ISCX}\\\textbf{2012}\end{tabular} & \begin{tabular}{@{}c@{}}\textbf{CIC}\\\textbf{2017}\end{tabular} & \begin{tabular}{@{}c@{}}\textbf{CSECIC}\\\textbf{2018}\end{tabular} & \begin{tabular}{@{}c@{}}\textbf{UNB}\\\textbf{201X}\end{tabular} \\ \midrule[\heavyrulewidth]
			\textbf{\ourtool} & 2241 & 0.9889 & 0.9966 & 0.9987 & 0.9946  \\ \midrule
			3LSTM & 1004889 & 0.9880 & 0.9968 & 0.9987 & 0.9943 \\\bottomrule[\heavyrulewidth]
		\end{tabular}
		\end{adjustbox}
	\end{threeparttable}
\end{table}

The results presented in Table \ref{tab:deepdefense_f1_comparison} show that \ourtool\ and 3LSTM are comparable in terms of F1 score across the range of test datasets. However, in terms of computation time, \ourtool\ 
outperforms 3LSTM in detection time. Specifically, as measured on the Intel Xeon server in these experiments, \ourtool\ can classify more than 55000 samples/sec on average, while 3LSTM barely reaches 1300 samples/sec on average (i.e., more than 40 times slower). Indeed, \ourtool's limited number of hidden units and trainable parameters contribute to a much lower computational complexity compared to 3LSTM.

As previously noted, there are a number of solutions in the literature that present performance results for the ISCX2012 and CIC2017 datasets. Notably, these works do not all specify whether the results presented are based on a validation dataset or a test dataset. For \ourtool{}, we reiterate that the results presented in this section are based on a test set of completely unseen data. 

\begin{table}[h!]
	\caption{Performance comparison with State-Of-The-Art approaches using the ISCX2012 dataset for \ac{ddos} detection.} 
	\label{tab:ISCXIDS2012_comparison}
	\small 
	\centering 
	\begin{threeparttable}
		\begin{tabular}{lccccc} \toprule[\heavyrulewidth]
			\textbf{Model}  & \textbf{ACC} & \textbf{FPR} & \textbf{PPV} & \textbf{TPR} & \textbf{F1} \\ \midrule[\heavyrulewidth]
			\textbf{\ourtool} & 0.9888 & 0.0179 & 0.9827 & 0.9952 & 0.9889  \\ \midrule
			\begin{tabular}{@{}l@{}}DeepDefense \\3LSTM \cite{7946998} \end{tabular}  & 0.9841 & N/A & 0.9834 & 0.9847 & 0.9840 \\ \midrule
			TR-IDS \cite{tr-ids}  & 0.9809 & 0.0040 & N/A & 0.9593 & N/A \\ \midrule
			E3ML \cite{8343104}  & N/A & N/A & N/A & 0.9474 & N/A \\ \bottomrule[\heavyrulewidth]
		\end{tabular}
	\end{threeparttable}
\end{table}

In Table \ref{tab:ISCXIDS2012_comparison}, we compare the performance of \ourtool\ against state-of-the-art works validated on ISCX2012. Table \ref{tab:ISCXIDS2012_comparison} also includes the performance of 3LSTM as reported in the DeepDefense paper \cite{7946998}. With respect to our version of 3LSTM, the scores are slightly lower, which we propose is due to the different \textit{pcap} preprocessing mechanisms used in the two implementations. This indicates a performance benefit when using the \ourtool{} preprocessing mechanism.

TR-IDS \cite{tr-ids} is an \ac{ids} which adopts a text-\ac{cnn} \cite{KimConvolutionalNN} to extract features from the payload of the network traffic. These features, along with a combination of 25 packet and flow-level attributes, are used for traffic classification by means of a Random Forest algorithm. Accuracy and \ac{tpr} of TR-IDS are above 0.99 for all the attack profiles available in ISCX2012 except the \ac{ddos} attack, for which the performance results are noticeably lower than \ourtool. 

E3ML \cite{8343104} uses 20 entropy-based traffic features and three \ac{ml} classifiers (a \ac{rnn}, a Multilayer Perceptron and an Alternating Decision Tree) to classify the traffic as normal or \ac{ddos}. Despite the complex architecture, the \ac{tpr} measured on ISCX2012 shows that E3ML is inclined to false negatives.

For the CIC2017 dataset, we present the performance comparison with state-of-the-art solutions in Table \ref{tab:CICIDS2017_comparison}.
 
\begin{table}[h!]
	\caption{Performance comparison with State-Of-The-Art approaches using the CIC2017 dataset for \ac{ddos} detection.} 
	\label{tab:CICIDS2017_comparison}
	\small 
	\centering 
	\begin{threeparttable}
		\begin{adjustbox}{width=0.49\textwidth}
		\begin{tabular}{lccccc} \toprule[\heavyrulewidth]
			\textbf{Model}  & \textbf{ACC} & \textbf{FPR} & \textbf{PPV} & \textbf{TPR} & \textbf{F1} \\ \midrule[\heavyrulewidth]
			\textbf{\ourtool} & 0.9967 & 0.0059 & 0.9939 & 0.9994 & 0.9966  \\ \midrule
			DeepGFL \cite{8599821}  & N/A & N/A & 0.7567 & 0.3024 & 0.4321 \\ \midrule
			MLP \cite{8666588}  & 0.8634 & N/A & 0.8847 & 0.8625 & 0.8735 \\ \midrule
			1D-CNN \cite{8666588}  & 0.9514 & N/A & 0.9814 & 0.9017 & 0.9399 \\ \midrule
			LSTM \cite{8666588}  & 0.9624 & N/A & 0.9844 & 0.8989 & 0.8959 \\ \midrule
			\begin{tabular}{@{}l@{}}1D-CNN +\\LSTM \cite{8666588}\end{tabular} & 0.9716 & N/A & 0.9741 & 0.9910 & 0.9825  \\ \bottomrule[\heavyrulewidth]
		\end{tabular}
		\end{adjustbox}
	\end{threeparttable}
\end{table}

DeepGFL \cite{8599821} is a framework designed to extract high-order traffic features from low-order features forming a hierarchical graph representation. To validate the proposed framework, the authors used the graph representation of the features to train two traffic classifiers, namely Decision Tree and Random Forest, and tested them on CIC2017. Although the precision scores on the several attack types are reasonably good (between 0.88 and 1 on any type of traffic profile except \ac{ddos}), the results presented in the paper reveal that the proposed approach is prone to false negatives, leading to very low F1 scores.

The authors of \cite{8666588} propose four different \ac{dl} models for \ac{ddos} attack detection in \ac{iot} networks. The models are built with combinations of LSTM, \ac{cnn} and fully connected layers. The input layer of all the models consists of 82 units, one for each flow-level feature available in CIC2017, while the output layer returns the probability of a given flow being part of a \ac{ddos} attack. The model 1D-CNN+LSTM produces good classification scores, while the others seem to suffer from high false negatives rates. 

To the best of our knowledge, no \ac{ddos} attack detection solutions validated on the CSECIC2018 dataset are available yet in the scientific literature.\\

\subsection{Discussion}
From the results presented and analysed in the previous sections, we can conclude that using packet-level attributes of network traffic is more effective, and results in higher classification accuracy, than using flow-level features or statistic information such as the entropy measure. This is not only proved by the evaluation results obtained with \ourtool\ and our implementation of DeepDefense (both based on packet-level attributes), but also by the high classification accuracy of TR-IDS, which combines flow-level features with packet attributes, including part of the payload. 

In contrast, E3ML, DeepGFL and most of the solutions proposed in \cite{8666588}, which all rely on flow-level features, seem to be more prone to false negatives, and hence to classify \ac{ddos} attacks as normal activity. The only exception is the model 1D-CNN+LSTM of \cite{8666588}, which produces a high \ac{tpr} by combining \ac{cnn} and \ac{rnn} layers. 

Furthermore, we highlight that \ourtool{} has not been tuned to the individual datasets but rather to the validation portion of a combined dataset, and still outperforms the state-of-the-art on totally unseen test data.

\section{Analysis}\label{sec:analysis}
We now present interpretation and explanation of the internal operations of \ourtool\ by way of proving that the model is learning the correct domain information. We do this by analysing the features used in the dataset and their activations in the model. To the best of our knowledge, this is the first application of a specific activation analysis to a \ac{cnn}-based \ac{ddos} detection method.

\subsection{Kernel activations}
This approach is inspired by a similar study \cite{jacovi-etal-2018-understanding} to interpret \acp{cnn} in the rather different domain of natural language processing.  However, the kernel activation analysis technique is transferable to our work.  As each kernel has the same width as the input matrix, it is possible to remove the classifier, push the \ac{ddos} flows through the convolutional layer and capture the resulting activations per kernel.  For each flow, we calculate the total activations per feature, which in the spatial input representation means per column, resulting in 11 values that map to the 11 features.  This is then repeated for all kernels, across all \ac{ddos} flows, with the final output being the total column-wise activation of each feature.  The intuition is that the higher a feature's activation when a positive sample i.e. a \ac{ddos} flow is seen, the more importance the \ac{cnn} attaches to that particular feature.  Conversely, the lower the activation, the lower the importance of the feature, and since our model uses the conventional rectified linear activation function, $ReLU(x) = max\{0,x\}$, this means that any negative activations become zero and hence have no impact on the Sigmoid classifier for detecting a DDoS attack. 

Summing these activations over all kernels is possible since they are of the same size and operate over the same spatial representations.  We analyse \ac{ddos} flows from the same UNB201X test set used in Sec. V-A. 

Table \ref{tab:col-activ} presents the ranking of the 11 features based on the post-$ReLU$ average column-wise feature activation sums, and highlights two features that activate our \ac{cnn} the most, across all of its kernels.  

\begin{table}[h!]
	\caption{Ranking of the total column-wise feature kernel activations for the UNB201X dataset}
	\label{tab:col-activ}
	\small 
	\centering 
	\renewcommand{\arraystretch}{1}
	\begin{tabular}{lc|lc} \toprule[\heavyrulewidth]
		\textbf{Feature} & \begin{tabular}{@{}c@{}}\textbf{Total Kernel}\\\textbf{Activation}\end{tabular} & \textbf{Feature} & \begin{tabular}{@{}c@{}}\textbf{Total Kernel}\\\textbf{Activation}\end{tabular} \\ 
		\midrule[\heavyrulewidth]
		Highest Layer  & 0.69540 & Time & 0.11108 \\\hline
		IP Flags  & 0.30337 & TCP Win Size & 0.09596 \\\hline
		TCP Flags  & 0.19693 & TCP Ack  & 0.00061 \\\hline
		TCP Len & 0.16874 & UDP Len  & 0.00000 \\\hline
		Protocols  & 0.14897 & ICMP Type  & 0.00000 \\ \hline
		Pkt Len & 0.14392 & & \\
		\bottomrule[\heavyrulewidth]
		
	\end{tabular}
\end{table}

\textbf{Highest Layer.}
We assert that the \ac{cnn} may be learning from the highest layer at which each \ac{ddos} flow operates.  Recall that highest layer links to the type of \ac{ddos} attack e.g. network, transport, or application layer attack. We propose that this information could be used to extend \ourtool{} to predict the specific type of \ac{ddos} attack taking place, and therefore, to contribute to selection of the appropriate protection mechanism. We would achieve the prediction by extending the dataset labeling, which we consider for future work. 

\textbf{IP Flags.} 
In our design, this attribute is a 16-bit integer value which includes three bits representing the flags \textit{Reserved Bit}, \textit{Don't Fragment} and \textit{More Fragments}, plus 13 bits for the \textit{Fragment offset} value, which is non-zero only if bit \textit{``Don't Fragment''} is unset. Unlike the \textit{IP fragmented flood} \ac{ddos} attacks, in which the \textit{IP flags} are manipulated to exploit the datagram fragmentation mechanisms, $99.99\%$ of \ac{ddos} packets in the \ac{unb} datasets present an \textit{IP flags} value of 0x4000, with only the \textit{``Don't Fragment''} bit set to $1$. A different distribution of \textit{IP flags} is observed in the \ac{unb} benign traffic, with the \textit{``Don't Fragment''} bit set to $1$ in about $92\%$ of the packets. Thus, the pattern of \textit{IP flags} is slightly different between attack and benign traffic, and we are confident that \ourtool\ is indeed learning their significance in \ac{ddos} classification, as evidenced by its 2nd place in our ranking.	

\subsection{Future Directions}
However, even given this activation analysis, there is no definitive list of features that exist for detecting \ac{ddos} attacks with which we can directly compare our results. Analysing the related work, we identify a wide range of both stateless and stateful features highlighted for their influence in a given detection model, which is not unexpected as the features of use vary depending on the attack traffic. This is highlighted by the 2014 study \cite{FIMU}, which concludes that different classes of attack have different properties, leading to the wide variance in features identified as salient for the attack detection. The authors also observe that the learning of patterns specific to the attack scenario would be more valuable than an effort to produce an attack-agnostic finite list of features. We, therefore, conclude from our analysis that \ourtool{} appears to be learning the importance of relevant features for \ac{ddos} detection, which gives us confidence in the prediction performance. 

Linked to this activation analysis, we highlight adversarial robustness as a key consideration for the deployment of ML-based \acp{ids}. As detailed in \cite{CoronaAdversarial:2013:AAA:2479999.2480270}, the two main attacks on \acp{ids} are during training via a poisoning attack (i.e. corruption of the training data), or in testing, when an evasion attack attempts to cause incorrect classification by making small perturbations to observed features. Our activation analysis is a first step in the investigation of the model behaviour in adversarial cases with the feature ranking in Table \ref{tab:col-activ} highlighting the features for perturbation for evasion attacks. Of course, the adversary model (goal, knowledge, and capability) dictates the potential for a successful attack. For example, the attacker would require full knowledge of the CNN and kernel activations, and have the ability to forge traffic within the network. The construction of defences robust to adversarial attacks is an open problem \cite{onevaluatingcarlini} and an aspect which we will further explore for \ourtool{}. 
\section{Use-case: DDoS detection at the edge}\label{sec:usecase}
Edge computing is an emerging paradigm adopted in a variety of contexts (e.g. fog computing \cite{Bonomi2014}, edge clouds \cite{6849256}), with the aim of improving the performance of applications with low-latency and high-bandwidth requirements. Edge computing complements centralised data centres with a large number of distributed nodes that provide computation services close to the sources of the data.

The proliferation of attacks leveraging unsecured \ac{iot} devices (e.g., the Mirai botnet \cite{mirai} and its variants) demonstrate the potential value in edge-based \ac{ddos} attack detection. Indeed, with edge nodes close to the \ac{iot} infrastructure, they can detect and block the \ac{ddos} traffic as soon as it leaves the compromised devices. However, in contrast to cloud high-performance servers, edge nodes cannot exploit sophisticated solutions against \ac{ddos} attacks, due to their limited computing and memory resources. Although recent research efforts have demonstrated that the mitigation of \ac{ddos} attacks is feasible even by means of commodity computers \cite{smartnic-ddos,Hoiland-Jorgensen:2018:EDP:3281411.3281443}, edge computing-based \ac{ddos} detection is still at an early stage.

In this section, we demonstrate that our \ac{ddos} detection solution can be deployed and effectively executed on resource-constrained devices, such as edge nodes or \ac{iot} gateways, by running \ourtool\ on an NVIDIA Jetson TX2 development board \cite{jetson-tx2}, equipped with a quad-core ARM Cortex-A57@2\,GHz CPU, 8\,GB of RAM and a 256-core Pascal@1300\,MHz \ac{gpu}. For the experiments, we used Tensorflow 1.9.0 with GPU support enabled by cuDNN, a GPU-accelerated library for deep neural networks \cite{cuDNN}.

\subsection{Detection}
In the first experiment, we analyse the applicability of our approach to online edge computing environments by estimating the prediction performance in terms of samples processed per second. As we are aware that edge nodes do not necessarily mount a \ac{gpu} device, we conduct the experiments with and without the \ac{gpu} support on the UNB201X test set and discuss the results.

We note that in an online system, our preprocessing tool presented in Section \ref{sec:pcap-processing} can be integrated into the server/edge device. The tool would process the live traffic collected from the NICs of the server/edge device, collecting the packet attributes, organising them into flows and, after a predefined time interval, $T$, pass the data structure to the CNN for inference. We acknowledge that the speed of this process will influence the overall system performance. However, as we have not focused on optimising our preprocessing tool, rather on optimising detection, its evaluation is left as future work. Instead, in these experiments, we load the \ac{unb} datasets from the hard disk rather than processing live traffic.

With respect to this, one relevant parameter is the batch size, which configures how many samples are processed by the \ac{cnn} in parallel at each iteration. Such a parameter influences the speed of the detection, as it determines the number of iterations  and, as a consequence, the number of memory reads required by the \ac{cnn} to process all the samples in the test set (or the samples collected in a time window, in the case of online detection).

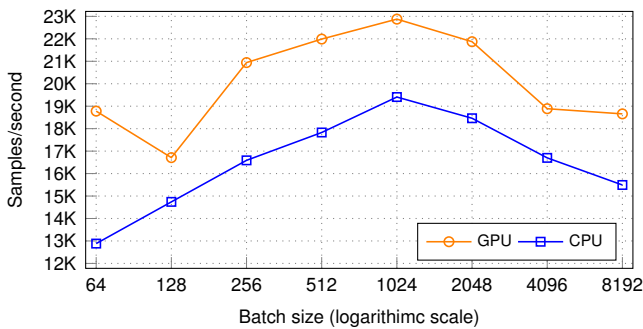
\begin{figure}[h!]
	\begin{tikzpicture}
	\begin{semilogxaxis}[  
	legend pos=south east,
	legend columns=2,
	height=5 cm,
	width=1\linewidth,
	grid = both,
	xlabel={Batch size (logarithimc scale)},
	ylabel={Samples/second},
	scaled y ticks=false,
	scaled x ticks=false,
	xmin=64,
	xmax=8192,
	xtick={64,128,256,512,1024,2048,4096,8192},
	xticklabels={64,128,256,512,1024,2048,4096,8192},
	xtick pos=left,
	ymin=12000, ymax=23000,
	ytick={12000,13000,14000,15000,16000,17000,18000,19000,20000,21000,22000,23000},
	yticklabels={12K,13K,14K,15K,16K,17K,18K,19K,20K,21K,22K,23K},
	ytick pos=left,
	enlargelimits=0.02,
	]
	\addplot [color=orange,style=semithick,mark=o,mark size=1.8] table[x index=0,y index=2] {\PredictionTX};
	\addplot [color=blue,style=semithick,mark=square,mark size=1.6] table[x index=0,y index=4] {\PredictionTX};
	\legend{GPU,CPU}
	\end{semilogxaxis}
	\end{tikzpicture}
	\caption{Inference performance on the NVIDIA Jetson TX2 board.}
	\label{fig:prediction_time}
\end{figure}

Fig. \ref{fig:prediction_time} shows the performance of \ourtool\ on the development board in terms of processed samples/second. As the shape of each sample is $[n,f]=[100,\features]$, i.e. each sample can contain the features of up to 100 packets, we can estimate that the maximum number of packets per second (pps) that the device can process without the \ac{gpu} and using a batch size of 1024 samples is approximately 1.9\,Mpps. As an example, the content of the UNB201X test set is 602,547 packets distributed over 22,416 samples, which represents a processing requirement of 500\,Kpps without the \ac{gpu}, and 600\,Kpps when the \ac{gpu} is enabled. This illustrates the ability to deploy \ourtool\ on a resource-constrained platform.

The second measurement regarding resource-constrained systems is the memory requirement to store all the samples collected over a time window. The memory occupancy per sample is 8,800 bytes, i.e. $100\cdot\features=1100$ floating point values of 8 bytes each. As per Fig. \ref{fig:prediction_time}, the \ac{cnn} can process around 23K samples/second with the help of the \ac{gpu} and using a batch size of 1024. To cope with such a processing speed, the device would require approximately 20\,GB RAM for a $t=100$ time window. However, this value greatly exceeds the typical amount of memory available on edge nodes, in general (e.g., 1\,GB on Raspberry Pi 3 \cite{raspberry}, 2\,GB on the ODROID-XU board \cite{8039523}), and on our device, in particular. Indeed, the memory resources of nodes can represent the real bottleneck in an edge computing scenario.

Therefore, assuming that our edge node is equipped with 1\,GB RAM, the maximum number of samples that can be stored in RAM is approximately 100K (without taking into account RAM used by the operating system and applications). We have calculated that this memory size would be sufficient for an attack such as the HTTP-based \ac{ddos} attack in the CSECIC2018 dataset, for which we measured approximately 30K samples on average over a 100\,s time window. For more aggressive attacks, however, a strategy to overcome the memory limitation would be to configure the \ac{cnn} model with lower values of $t$ and $n$. For instance, setting the value of both parameters to 10 can reduce the memory requirement by a factor of 100, with a low cost in detection accuracy (F1 score 0.9928 on the UNB201X test set, compared to the highest score obtained with $t=n=100$, i.e. 0.9946). The dynamic configuration of the model itself is out of scope of this work.

The measurements based on our test datasets demonstrate that the \ourtool\ \ac{cnn} is usable on a resource-constrained platform both with respect to processing and memory requirements. These results are promising for effective deployment of \ourtool\ in a variety of edge computing scenarios, including those where the nodes execute latency-sensitive services. A major challenge in this regard is balancing between resource usage of \ourtool\ (including traffic collection and preprocessing) and detection accuracy, i.e. ensuring the required level of protection against \ac{ddos} attacks without causing delays to the services. A deep study of this trade-off is out of scope of this paper and is reserved for future work.

\subsection{Training time}
In a real-world scenario, the \ac{cnn} model will require re-training with new samples of benign and malicious traffic to update all the weights and biases. In edge computing environments, the traditional approach is to send large amounts of data from edge nodes to remote facilities such as private or commercial datacentres. However, this can result in high end-to-end latency and bandwidth usage. In addition, it may raise security concerns, as it requires trust in a third-party entity (in the case of commercial cloud services) regarding the preservation of data confidentiality and integrity. 

A solution to this issue is to execute the re-training task locally on the edge nodes. In this case, the main challenge is to control the total training time, as this time determines how long the node remains exposed to new \ac{ddos} attacks before the detection model can leverage the updated parameters.

To demonstrate the suitability of our model for this situation, we have measured the convergence training time of \ourtool\ on the development board using the UNB201X training and validation sets with and without the \ac{gpu} support. We have experimented by following the learning procedure described in Sec. \ref{sec:learning}, thus with a training termination criterion based on the loss value measured on the validation set. The results are presented in Table \ref{tab:training_tx} along with the performance obtained on the server used for the study in Sec. \ref{sec:tuning}. 

\begin{table}[h!]
	\caption{Training convergence time.} 
	\label{tab:training_tx}
	\small 
	\centering 
	\begin{threeparttable}
		\begin{tabular}{lcc} \toprule[\heavyrulewidth]
			\textbf{Setup}  & \begin{tabular}{@{}c@{}}\textbf{Time/epoch} \\\textbf{(sec)} \end{tabular} & \begin{tabular}{@{}c@{}}\textbf{Convergence} \\\textbf{time (sec)} \end{tabular} \\ \midrule[\heavyrulewidth]
			\textbf{\ourtool}\ Server  & 10.2 & 1880 \\ \midrule
			\textbf{\ourtool}\ Dev. board (\ac{gpu}) & 25.8 & 4500  \\ \midrule
			\textbf{\ourtool}\ Dev. board (CPU) & 40.5 & 7450   \\ \midrule
			3LSTM Dev. board (\ac{gpu}) & 1070 & $>$90000   \\ \bottomrule[\heavyrulewidth]
		\end{tabular}
	\end{threeparttable}
\end{table}

As shown in Table \ref{tab:training_tx}, the \ac{cnn} training time on the development board without using the \ac{gpu} is around 2 hours (184 epochs). This is approximately 4 times slower than training on the server, but clearly outperforms the training time of our implementation of DeepDefense 3LSTM, which we measured at more than 1000 sec/epoch \textit{with} the \ac{gpu} (i.e., 40 times slower than \ourtool{} under the same testing conditions).  

In application scenarios where a faster convergence is required, the time can be further reduced by either terminating the training process early after a pre-defined number of epochs, or limiting the size of the training/validation sets. As adopting one or both of such strategies can result in a lower detection accuracy, the challenge in such scenarios is finding the trade-off between convergence time and detection accuracy that meets the application requirements.
\section{Conclusions}\label{sec:conclusions}

The challenge of \ac{ddos} attacks continues to undermine the availability of networks globally. In this work, we have presented a \ac{cnn}-based \ac{ddos} detection architecture. Our design has targeted a practical, lightweight implementation with low processing overhead and attack detection time. The benefit of the \ac{cnn} model is to remove threshold configuration as required by statistical detection approaches, and reduce feature engineering and the reliance on human experts required by alternative \ac{ml} techniques. This enables practical deployment. 

In contrast to existing solutions, our unique traffic pre-processing mechanism acknowledges how traffic flows across network devices and is designed to present network traffic to the \ac{cnn} model for online \ac{ddos} attack detection. Our evaluation results demonstrate that \ourtool\ matches the existing state-of-the-art performance. However, distinct from existing work, we demonstrate consistent detection results across a range of datasets, demonstrating the stability of our solution. Furthermore, our evaluation on a resource-constrained device demonstrates the suitability of our model for deployment in resource-constrained environments. Specifically, we demonstrate a 40x improvement in processing time over similar state-of-the-art solutions. Finally, we have also presented an activation analysis to explain how \ourtool{} learns to detect \ac{ddos} traffic, which is lacking in existing works.

\section*{Acknowledgment}
This work has received funding from the European Union's Horizon 2020 Research  and Innovation Programme under grant agreement no. 815141 (DECENTER project).

\bibliographystyle{IEEEtran} 
\bibliography{bibliography}

\end{document}